\documentclass[11pt,twoside]{article}

\usepackage{newpasp}
\usepackage{epsf}
\usepackage{epsfig}
\usepackage{rotating}
\usepackage{makeidx}
\usepackage{graphicx}

\def\edcomment#1{\iffalse\marginpar{\raggedright\sl#1\/}\else\relax\fi}
\reversemarginpar

\begin{document}
\title{Asymptotic Giant Branch Stars in Interacting Galaxy Globular 
       Clusters}
\author{M. Mouhcine \& A. Lan\c{c}on}
\affil{Observatoire de Strasbourg, 11 rue de l'Universit\'e, 67000 Strasbourg,
France}

\begin{abstract}
We present new modelling of the spectrophotometric properties of intermediate
age stellar populations in the near-infrared (NIR). We take into account the 
evolutionary and spectroscopic properties of the Asymptotic Giant Branch (AGB)
stars which dominate the integrated NIR emission of those populations.
The predictions of spectrophotometric narrow-band molecular indices
require the use of an effective temperature -- colour (or spectrum)
calibration for AGB stars; synthetic indices show a strong dependence 
upon this calibration.

Preliminary results of NIR observations of a supermassive 
500\,Myr old stellar cluster in the prototypical merger remnant 
NGC 7252 are presented. The spectra are indeed consistent with 
intermediate age models dominated by light from AGB stars, a significant 
fraction of which may be carbon rich. 
Implications for the stellar inputs used in the modelling 
of intermediate age stellar populations and their NIR emission are discussed. 
\end{abstract}

\keywords{AGB stars -- carbon stars -- mergers}

\section{Introduction}
Space and ground based observations have shown that the star formation in
starburst regions appears to be biased toward compact clusters. This process
seems to be common to all galaxy types: merging galaxies (NGC 3597: Lutz 1991,
NGC 1275: Holtzman et al. 1992, NGC 4038/39 Whitmore et al. 1999), starburst 
galaxies (NGC 1705: Meurer et al. 1992, M82: Gallagher \& Smith 1999, Smith 
\& Gallagher, this volume), dwarf galaxies 
(Henize 2-10: Conti \& Vacca 1994, ESO338-IG04, \"Ostlin et al. 1998) and even 
barred galaxies (NGC 1097: Barth et al. 1995). Colours, luminosities and 
spatial properties of those clusters are consistent with their being globular 
clusters with effective radii of a few parsecs (Schweizer 1999).

It is established that the AGB stars are the principal contributors to the 
near infrared (NIR) luminosity of intermediate age stellar 
populations\footnote{We define intermediate age populations as those having 
ages between 10$^8$ and 10$^9$\,yr.} (Persson et al. 1983). 
The AGB evolutionary phase translates into
major changes of the HR diagram properties of stellar populations over small 
time scales. 
Traditionally, the spectroscopic diagnostic tools that are used in the NIR for 
age dating are principally based on CO bands (2.3$\,\mu$m) or broad-band 
colours. The accuracy of those methods is questionable. After the disappearance 
of massive supergiants that display very deep CO bands, it becomes hard to 
discriminate between AGB stellar populations and the red giant branch (RGB) 
stellar populations on the basis of CO bands only. 
Changes in the NIR colours are ambiguous because of degeneracies with 
the effects of metallicity and extinction.
Evidence for strong colour changes due to the AGB stars is found principally 
in observations of the Magellanic Cloud globular clusters. However, the 
spectrophotometry of those clusters is affected by large stochastic
fluctuations due to the small numbers of luminous cool stars. The 
interpretation of those observations is not very secure and larger
samples (not available in the LMC) are required.

Recently, Lan\c{c}on et al. (1999) have proposed new diagnostic tools to 
identify the intermediate age populations using other molecular absorption 
bands than the CO bands. They distinguish between the oxygen rich and carbon 
rich AGB stars contributing to the NIR light of post-starburst populations 
based on the spectroscopic features of the AGB stars that are absent in the 
RGB star spectra.  

\section{Intermediate Age Stellar Populations in the NIR: Modelling and 
Uncertainties}

The synthesis of stellar populations and their emission is initially
based on the code {\sc P\'egase} (Fioc \& Rocca-Volmerange 1997,
Fioc 1997), with considerable adjustments to allow for changes in all 
the AGB-related inputs. In this section we present the main features of the 
modified models.

>From the main sequence and up to the Early-AGB phase we use the tracks of 
the Padova group (Bressan et al. 1993, Fagotto et al. 1994).
The physical processes controlling the detail of the evolution along the
thermally pulsing AGB (TP-AGB) are complex (e.g. Iben \& Renzini 1983).
Models are still affected by a large degree of uncertainty and 
are still a matter of debate (Habing 1996, Olofsson 1999). 
Full calculations of AGB stars (internal structure, evolution,
pulsation-stimulated mass loss) for a large grid of initial masses and 
metallicities are still missing in the literature (see  
Vassiliadis \& Wood 1993 for 6 initial masses at solar and LMC metallicities). 
In this context, the so-called synthetic evolution models 
based on analytical formula derived from full numerical calculations 
provide an ideal tool to study evolutionary and chemical aspects of 
TP-AGB stars (Iben \& Truran 1978, Renzini \& Voli 1981, 
Groenewegen \& de Jong 1993, Marigo et al. 1996, 
Wagenhuber \& Groenewegen 1998). They make it possible to model, 
in a simple but relatively complete way, the complex interplay between the 
many relevant physical processes that together rule TP-AGB life.
They allow us to cover a large grid of parameters and to reproduce 
the basic observational constraints obtained mostly from Magellanic Cloud
and Milky Way observations. 
We have chosen to model TP-AGB evolution with such a synthetic code, 
including: (1) Mass loss, with winds
reaching the superwind regime with final rates as high as 
$10^{-4}\,M_{\odot}\,yr^{-1}$ (we investigate the prescriptions of
various authors, but do not currently include 
any explicit metallicity dependence; the implicit dependance through 
effective temperature and luminosity remains). 
Winds play a dominant role in determining the lifetime and the extent 
of nuclear processing in TP-AGB stars. 
(2) The third dredge-up phenomenon which is responsible for the formation 
of carbon-rich stars characterized by very red colours (especially in J-K) 
and peculiar spectral features. Dredge-up also affects the core 
mass\,--\,luminosity relation. 
(3) Hot bottom burning, which affects the evolution of massive stars 
($M_{ZAMS}\,\ge\,3-4\,M_{\odot}$ for solar metallicity) preventing the 
formation of carbon-rich stars from the progenitors of massive core stars 
and drastically reducing the lifetime of those stars, which reduce their 
contribution to the integrated (particularly the NIR) light of a population. 
We assume the TP-AGB phase to be complete either by the total ejection of the 
envelope, or when the core mass has reached the Chandrasekhar limit 
($\sim\,1.4\,M_{\odot}$). In contrast to other synthesis models for
the  spectrophotometric evolution of stellar populations, the evolution of the 
chemical types of the stars along the TP-AGB is followed here as a result of 
the competition between all those processes. 
The values of the free parameters of the synthetic AGB star evolution models
are, as a starting point, taken from the literature 
(Groenewegen \& de Jong 1993, Marigo et al. 1996). 

In addition to the thermal pulses, occurring on a time scale of the order 
of $10^4\,-\,10^5$ years, the AGB stars exhibit another type of variation:
the Mira-like pulsation, with a time scale of the order of 
$10^{2}-10^{3}$ days. Those pulsations lead to large effective 
temperature shifts and have strong effects on the NIR spectral features of 
upper AGB stars. The latter display stronger and deeper molecular bands than 
any static giant or supergiant. The relevant molecules are
H$_{2}$O, TiO and VO for oxygen-rich stars, 
CN and C$_{2}$ for carbon-rich stars. 
Those spectral particularities give us the possibility to 
recognise TP-AGB stars from RGB stars. 
They are the basis of our strategy to look for 
intermediate age populations, and for dating the stellar populations 
(Mouhcine \& Lan\c{c}on 1999, Lan\c{c}on 1999). 
Introducing the mechanical Mira-like pulsations explicitly in the calculations 
of the evolutionary tracks is impossible due to the different time scales of 
the two effects. This effect is included in our calculations via a new stellar 
spectral library. Although the libraries of stellar spectra available 
in the literature include various spectral types, they do not include variable 
stars yet. To overcome this problem, we have constructed a new extension
designed specially for our applications. The 
library relies on suitable averaging procedures over a large data set of NIR 
spectra of cool AGB stars with various periods (Lan\c{c}on \& Wood, submitted).
The innovative aspect of this library is that it includes spectra of objects 
at different evolutionary stages along the AGB: M stars, carbon rich stars, 
OH/IR stars and C/IR stars.
To assign effective temperatures to the stars in our spectral library 
is a difficult task. To be conservative, we have used two different effective 
temperature scales for the variable AGB stars. The first one is based on
a large grid of static giant atmosphere models (Plez 1999) 
and fits to the whole spectral energy distribution of instantaneous stellar 
spectra without special care to the special spectral signatures of variable 
stars. The second one is established using angular diameter measurements and IR 
photometry of late-type stars for a mixed sample of mostly Mira-type 
stars by Feast (1996). The second scale is steeper than the 
first one. This means that when the first scale is used to relate the 
theoretical plane to the observational plane,
the very late type spectra (spectra with very pronounced signatures) are 
assigned a relatively high temperature and get used frequently in comparison 
to the case when Feast's temperature scale is used. This effect is clearly
shown in Fig.\,1 where the temporal evolution of two molecular indices is 
plotted.
\begin{figure}
\includegraphics[clip=,width=0.49\textwidth]{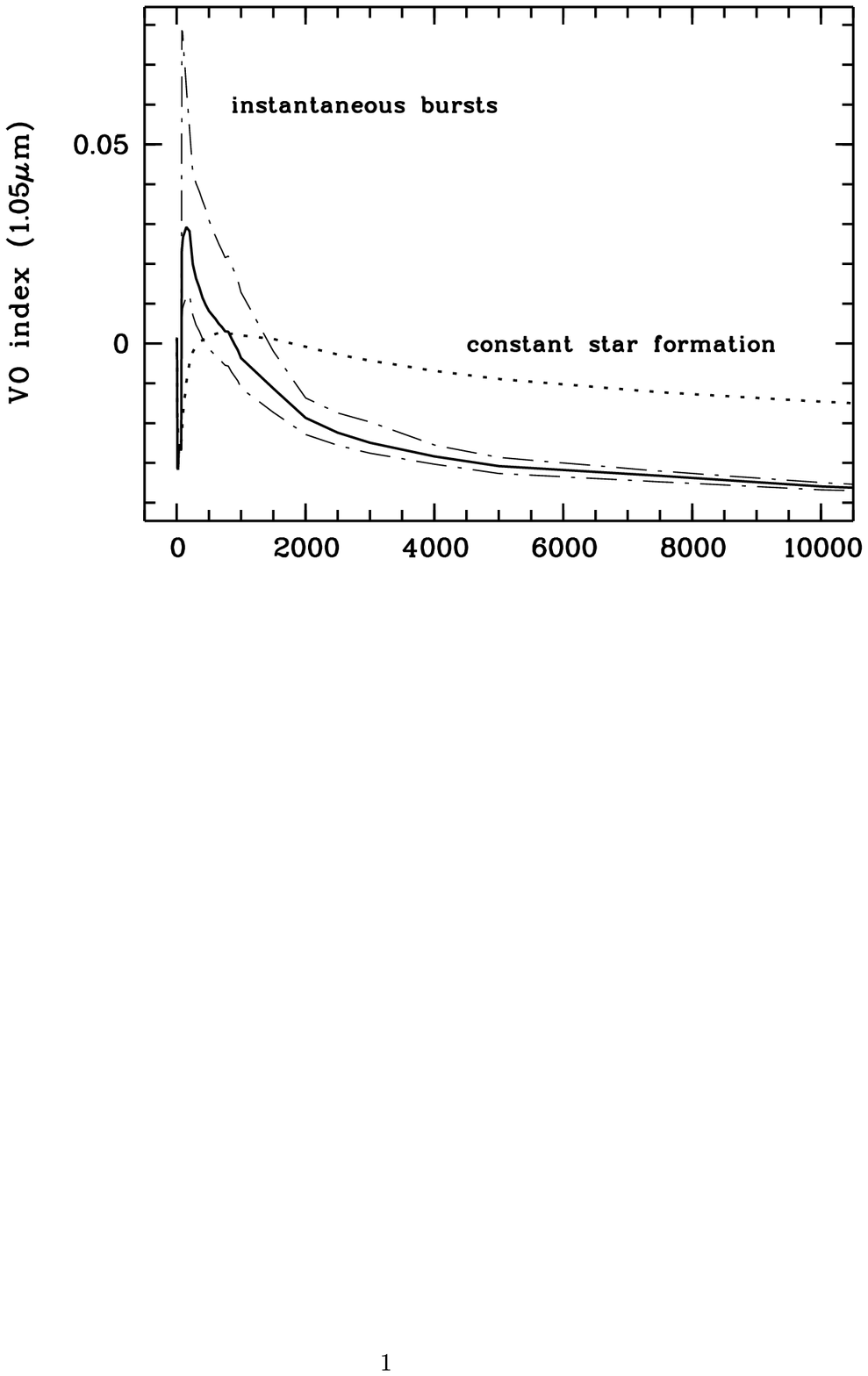}
\includegraphics[clip=,width=0.49\textwidth]{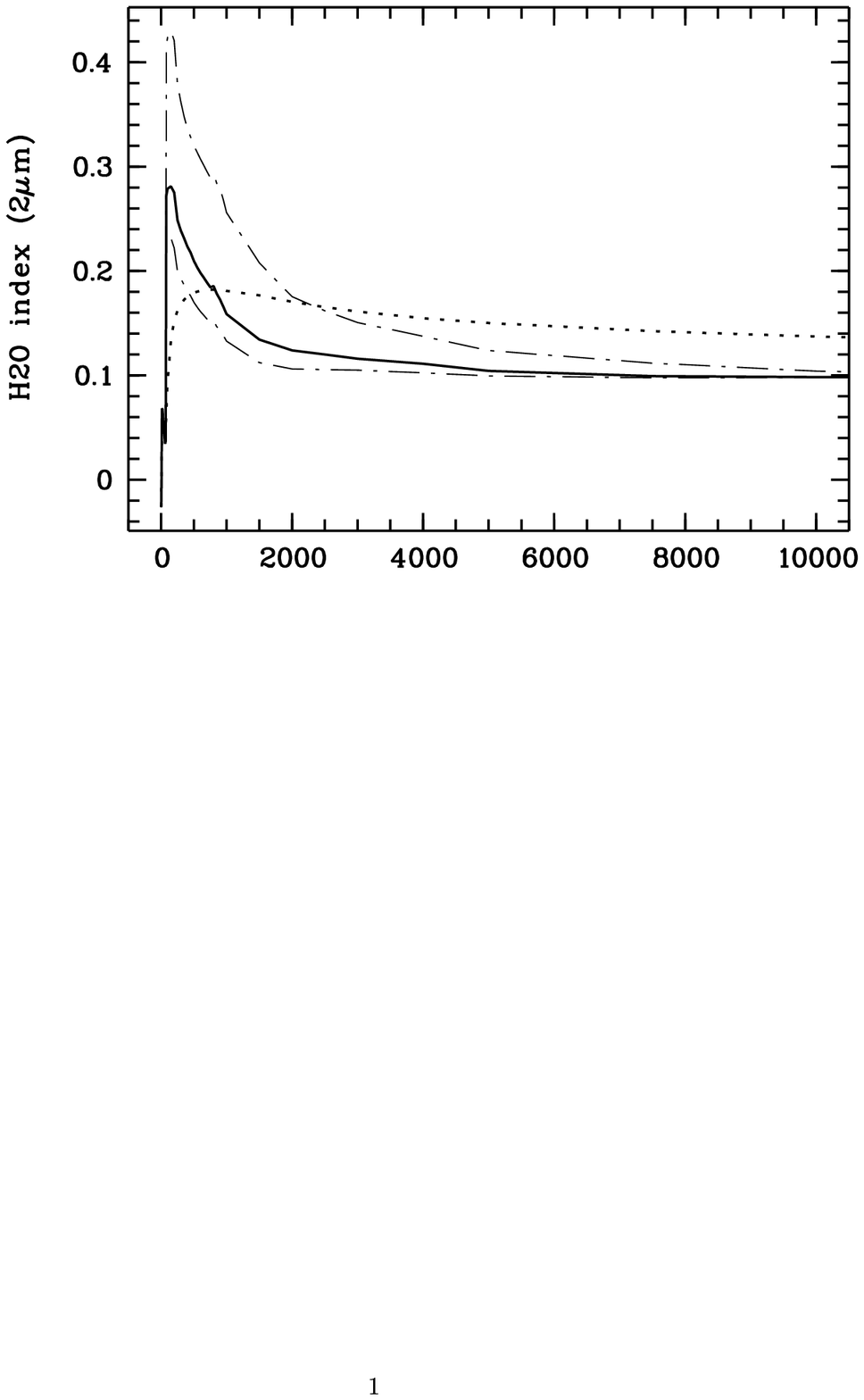}
\vspace{-1.7mm}

\tiny
~\hfill \ \ Time (Myr) \hfill ~ \hfill Time (Myr) \hfill ~
\normalsize
\caption{Evolution of selected molecular indices in the NIR for solar 
metallicity for an instantaneous burst and for constant star formation 
with the assumption that all AGB stars remain oxygen-rich. Dot-dashed
lines: the two extreme T$_{\mathrm eff}$ scales described in the text; 
solid and dotted: an intermediate scale.}
\end{figure}
\section{NGC 7252 Massive Cluster}

One of the best systems where to study supermassive clusters is the 
prototypical merger remnant NGC 7252. Using ground based observations, 
Schweizer (1982) identified many very luminous point-like sources in this galaxy.
Whitmore et al. (1993) and Miller et al. (1997) performed {\it Hubble 
Space Telescope} UBVI observations and confirmed these sources to be globular 
clusters. They estimated that the most luminous (massive) ones have ages 
between 0.1 and 1 Gyr and masses of up to $10^{8}\,M_{\odot}$.
Such systems not only represent an excellent laboratory to study the
impact of galaxy interactions on the star and cluster formation; to us,
the massive clusters are ideal systems to test the theoretical predictions for 
the evolution of spectrophotometric features dominated by rare
bright stars, which suffer from stochastic fluctuations in smaller 
populations (e.g. Lan\c{c}on \& Mouhcine, this volume
\typeout{missing - pageref Lancon - Mouhcine paper}). 
The general advantage of observing globular clusters comes from the fact 
that they contain (in general) a single stellar population, homogeneous 
in age and chemical composition.

In order to investigate the above questions, we have selected the most luminous 
cluster of NGC 7252, cluster W3.
Schweizer \& Seitzer (1998, hereafter SS98) have recently obtained
ultraviolet-to-visual spectra of W3. The most striking features are the
dominating strong Balmer absorption lines and the blue continuum, indicative
of A-type main-sequence stars. The spectra do not show any strong emission
lines, indicating that the system is not very young (age higher than 0.1 Gyr).
Based on comparisons between the observed and modelled spectra,
they derive for this cluster an age of about 0.5 Gyr and 
near-solar metallicity.

We have observed W3 with the infrared imager and spectrometer  
SOFI on the ESO New Technology Telescope in August 1999. SOFI has 
a field-of-view of $\sim 5'\!\times\!5'$ and a pixel size of $0.292''$.
The total spectroscopic exposure time was about 2.5 hours both in the blue 
wavelength range (0.9 to 1.61\,$\mu$m) and in the red wavelength range 
(1.6 to 2.5\,$\mu$m). Our spectra have a resolution of R$\simeq$1000 and a 
signal-to-noise ratio in the range of 30-35 (note that the seeing rarely
dropped below $1''$). The observations were reduced 
following standard procedures using IRAF software.

To fit the NIR observed W3 spectra, we have used an early version of
the population synthesis models presented in Sect.\,2, with the assumptions of 
instantaneous star formation and a standard Salpeter IMF extending from 0.1
to 120\,M$_{\odot}$.  We have adopted the age and the metallicity derived 
by SS98 and have tested the two effective temperature scales for AGB star 
spectra described in Sect.\,2. 

The comparison between the models and the W3 data strongly argues against 
a dominant contribution of the O-rich TP-AGB stars with the deepest
molecular features, i.e. against the warmer temperature scale based on 
static models. The steeper, cooler scale provides satisfactory 
agreement with the data.

Although the NGC\,7252 clusters are said to have a quasi-solar metallicity, 
we have compared the W3  data with a model at LMC metallicity, in which,
as a consequence of Groenewegen \& de Jong's (1993) relative C-rich and 
O-rich TP-AGB lifetimes, carbon stars dominate at 0.5\,Gyr. The agreement
with the shape of the features is better than with the solar metallicity
models, in which C-stars currently provide an insignificant flux contribution.
These may be the first direct observations of carbon stars at more than
50\,Mpc. In any case, it confirms that the inclusion of C-star formation
processes in spectrophotometric evolution models, as discussed 
in Sect.\,2, is relevant. More detail will be given in a forthcoming 
paper.

\section{Conclusions}
The use of massive globular clusters in merging galaxy remnants gives us
the opportunity to overcome the stochastic fluctuations affecting the NIR 
spectrophotometric properties of the intermediate age population due to the 
small number of bright AGB stars and to test, both qualitatively and 
quantitatively, the population synthesis models. We have observed 
the most massive cluster of NGC 7252 in the NIR. These observations 
allow us to report the first observations of a stellar population dominated 
by AGB stars beyond the Local Group. 

Comparisons with our model cluster spectra lead us to argue that the use of an 
effective temperature scale calibrated on static giant branch stars for TP-AGB 
stars is, at best, questionable. The observed spectrum is more consistent with 
model cluster spectra calculated using an effective temperature scale 
calibrated on LPV stars. The good fit obtained when a significant
fraction of the 500\,Myr old TP-AGB stars are carbon rich shows 
the importance of taking surface abundance evolution into account 
in models for the spectroscopic evolution of intermediate age 
populations.

\acknowledgements

It is a pleasure to acknowledge the fruitful collaboration with M.\,Fioc,
M.\,Groenewegen, C.Leitherer and D.\,Silva at various steps of this
work.

\begin{question}{B. Miller}
If spectra are unavailable, are JHK integral magnitudes 
also useful discriminators of your models? Are there other useful NIR 
bandpasses? How small do the uncertainties in the colors need to be?
\end{question}

\begin{answer}{M. Mouhcine}
Some near-IR colours, in particularly H-K, do get particularly
red in intermediate age populations in our models; however, there are dangerous
degeneracies with effects of dust and metallicity, for example
in the J-H versus H-K plane. Being able 
to determine these independently is at least as important as
good photometric accuracy. How red the colours become also 
strongly depends on the T$_{\rm eff}$ scale adopted for 
the TP-AGB star spectra; we believe that calibrations on more extended
cluster samples than just those of the LMC are needed before NIR colours
become a reliable age dating tool. Searching for spectrophotometric features
due to stellar pulsation has the advantage of avoiding ambiguities.
\end{answer}

\begin{question}{G. Meurer}
Maybe narrow band filters could be used in the IR to isolate spectral 
features, and thus observe many clusters at once.
\end{question}

\begin{answer}{M. Mouhcine}
This is exactly the observational strategy that we have 
proposed to discriminate between the intermediate-age populations dominated 
by the AGB stars and the other stellar populations (Lan\c{c}on et al. 1998). 
It is based on observations on different narrow bands. But as shown in the 
graph, the evolution of the molecular indices depends strongly on 
different stellar parameters and, again, must be 
calibrated using simple burst populations before it can be used in more 
complex stellar populations. In addition, since the relevant features
are rather close to telluric absorption bands and since good relative
photometry is needed, we still tend to prefer low resolution, broad
wavelength coverage spectroscopic methods at the moment.   
\end{answer}
\end{document}